\documentclass[12pt]{article} 

\pdfoutput=1
\usepackage{amsmath}
\usepackage{graphicx}
\usepackage{multirow}
\usepackage{times}
\usepackage{wrapfig}
\usepackage{color}
\usepackage{lscape}
\usepackage{sidecap}
\usepackage{fltpage}
\usepackage{tocloft}
\usepackage{setspace}
\usepackage{hyperref}
\usepackage[normalem]{ulem}

\def\letter#1{(\textbf{#1})}

\title{Flavor network and the principles of food pairing} 

\author{Yong-Yeol Ahn,$^{1,2,3\dagger}$ Sebastian E.~Ahnert,$^{1,4\dagger\ast}$ James P.~Bagrow,$^{1,2}$ \\
Albert-L\'aszl\'o Barab\'asi$^{1,2\ast}$\\
\\
\normalsize{$^{1}$Center for Complex Network Research, Department of Physics}\\
\normalsize{Northeastern University, Boston, MA 02115}\\
\normalsize{$^{2}$Center for Cancer Systems Biology}\\
\normalsize{Dana-Farber Cancer Institute, Harvard University, Boston, MA 02115}\\
\normalsize{$^{3}$School of Informatics and Computing}\\
\normalsize{Indiana University, Bloomington, IN 47408}\\
\normalsize{$^{4}$Theory of Condensed Matter, Cavendish Laboratory}\\
\normalsize{University of Cambridge, Cambridge CB3 0HE, UK}\\
\\
\normalsize{$^\dagger$ These authors contributed equally to this work.}\\
\normalsize{$^\ast$ To whom correspondence should be addressed.}\\ \normalsize{E-mail: sea31@cam.ac.uk (S.E.A.); alb@neu.edu (A.L.B.)}
}

\date{}

\begin{document} 
\baselineskip24pt 
\maketitle

\begin{abstract}

The cultural diversity of culinary practice, as illustrated by the variety of
regional cuisines, raises the question of whether there are any general
patterns that determine the ingredient combinations used in food today or
principles that transcend individual tastes and recipes. We introduce a flavor
network that captures the flavor compounds shared by culinary ingredients.
Western cuisines show a tendency to use ingredient pairs that share many flavor
compounds, supporting the so-called food pairing hypothesis. By contrast, East
Asian cuisines tend to avoid compound sharing ingredients. Given the increasing
availability of information on food preparation, our data-driven investigation
opens new avenues towards a systematic understanding of culinary practice.

\end{abstract} 

As omnivores, humans have historically faced the difficult task of identifying
and gathering food that satisfies nutritional needs while avoiding foodborne
illnesses~\cite{rozin1976}. This process has contributed to the current diet of
humans, which is influenced by factors ranging from an evolved preference for
sugar and fat to palatability, nutritional value, culture, ease of production,
and climate~\cite{rozin1976, pfaffman1975, garcia1975, drewnowski1983,
Diamond:1997, billing1998, sherman2001, goodtoeat, foodculture}.  The
relatively small number of recipes in use ($\sim 10^6$, e.g.
http://cookpad.com) compared to the enormous number of potential recipes ($>
10^{15}$, see Supplementary Information Sec S1.2), together with the frequent
recurrence of particular combinations in various regional cuisines, indicates
that we are exploiting but a tiny fraction of the potential combinations.
Although this pattern itself can be explained by a simple evolutionary
model~\cite{kinouchi_culinaryevol_2008} or data-driven
approaches~\cite{teng_recipe_2011}, a fundamental question still remains: are
there any quantifiable and reproducible principles behind our choice of certain
ingredient combinations and avoidance of others?

Although many factors such as colors, texture, temperature, and sound play an
important role in food sensation~\cite{this_molecular_gastronomy, johnson1982,
shankar2010, zampini_potatochips_2005}, palatability is largely determined by
flavor, representing a group of sensations including odors (due to molecules
that can bind olfactory receptors), tastes (due to molecules that stimulate
taste buds), and freshness or pungency (trigeminal senses)~\cite{breslin1995}.
Therefore, the flavor compound (chemical) profile of the culinary ingredients
is a natural starting point for a systematic search for principles that might
underlie our choice of acceptable ingredient combinations. 

A hypothesis, which over the past decade has received attention among some
chefs and food scientists, states that ingredients sharing flavor compounds are
more likely to taste well together than ingredients that do
not~\cite{blumenthal}.  This food pairing hypothesis has been used to search
for novel ingredient combinations and has prompted, for example, some
contemporary restaurants to combine white chocolate and caviar, as they share
\emph{trimethylamine} and other flavor compounds, or chocolate and blue cheese
that share at least 73 flavor compounds.  As we search for evidence supporting
(or refuting) any `rules' that may underlie our recipes, we must bear in mind
that the scientific analysis of any art, including the art of cooking, is
unlikely to be capable of explaining every aspect of the artistic creativity
involved.  Furthermore, there are many ingredients whose main role in a recipe
may not be only flavoring but something else as well (e.g.  eggs' role to
ensure mechanical stability or paprika's role to add vivid colors). Finally,
the flavor of a dish owes as much to the mode of preparation as to the choice
of particular ingredients~\cite{this_molecular_gastronomy, this:2009,
mcgee_cooking}. However, our hypothesis is that given the large number of
recipes we use in our analysis (56,498), such confounding factors can be
systematically filtered out, allowing for the discovery of patterns that may
transcend specific dishes or ingredients.

Here we introduce a network-based approach to explore the impact of flavor
compounds on ingredient combinations. Efforts by food chemists to identify the
flavor compounds contained in most culinary ingredients allows us to link each
ingredient to 51 flavor compounds on average~\cite{fenaroli}\footnote{While
finalizing this manuscript, an updated edition (6th Ed.) of \emph{Fenaroli's
handbook of flavor ingredients} has been released.}.  We build a bipartite
network~\cite{newman_structure_2006, caldarelli_scale-free_2007, doro_book,
barabasi02, newman_review_2003, dorogovtsev_critical_2008} consisting of two
different types of nodes: (i) 381 ingredients used in recipes throughout the
world, and (ii) 1,021 flavor compounds that are known to contribute to the
flavor of each of these ingredients (Fig.~1A). A projection of this bipartite
network is the \emph{flavor network} in which two nodes (ingredients) are
connected if they share at least one flavor compound (Fig.~1B). The weight of
each link represents the number of shared flavor compounds, turning the flavor
network into a weighted
network~\cite{barrat_weighted,caldarelli_scale-free_2007,doro_book}. While the
compound concentration in each ingredient and the detection threshold of each
compound should ideally be taken into account, the lack of systematic data
prevents us from exploring their impact (see Sec S1.1.2 on data limitations).

Since several flavor compounds are shared by a large number of ingredients, the
resulting flavor network is too dense for direct visualization (average degree
$\left< k \right> \simeq 214$). We therefore use a backbone extraction
method~\cite{serrano2009,shlee_google} to identify the statistically
significant links for each ingredient given the sum of weights characterizing
the particular node (Fig.~2), see SI for details).  Not surprisingly, each
module in the network corresponds to a distinct food class such as meats (red)
or fruits (yellow). The links between modules inform us of the flavor compounds
that hold different classes of foods together. For instance, fruits and dairy
products are close to alcoholic drinks, and mushrooms appear isolated, as they
share a statistically significant number of flavor compounds only with other
mushrooms.

The flavor network allows us to reformulate the food pairing hypothesis as a
topological property: do we more frequently use ingredient pairs that are
strongly linked in the flavor network or do we avoid them? To test this
hypothesis we need data on ingredient combinations preferred by humans,
information readily available in the current body of recipes. For generality,
we used 56,498 recipes provided by two American repositories
(\emph{epicurious.com} and \emph{allrecipes.com}) and to avoid a distinctly
Western interpretation of the world's cuisine, we also used a Korean repository
(\emph{menupan.com}) (Fig.~1). The recipes are grouped into geographically
distinct cuisines (North American, Western European, Southern European, Latin
American, and East Asian; see Table~S2).  The average number of ingredients
used in a recipe is around eight, and the overall distribution is bounded
(Fig.~1C), indicating that recipes with a very large or very small number of
ingredients are rare. By contrast, the popularity of specific ingredients
varies over four orders of magnitude, documenting huge differences in how
frequently various ingredients are used in recipes (Fig.~1D), as observed
in~\cite{kinouchi_culinaryevol_2008}.  For example, jasmine tea, Jamaican rum,
and 14 other ingredients are each found in only a single recipe (see SI S1.2),
but egg appears in as many as 20,951, more than one third of all recipes.

\section*{Results}

Figure~3D indicates that North American and Western European cuisines exhibit a
statistically significant tendency towards recipes whose ingredients share
flavor compounds. By contrast, East Asian and Southern European cuisines avoid
recipes whose ingredients share flavor compounds (see Fig.~3D for the
$Z$-score, capturing the statistical significance of $\Delta N_s$).  The
systematic difference between the East Asian and the North American recipes is
particularly clear if we inspect the $P(N^\mathrm{rand}_s)$ distribution of the
randomized recipe dataset, compared to the observed number of shared compounds
characterizing the two cuisines, $ N_s $.  This distribution reveals that North
American dishes use far more compound-sharing pairs than expected by chance
(Fig.~3E), and the East Asian dishes far fewer (Fig.~3F). Finally, we
generalize the food pairing hypothesis by exploring if ingredient pairs sharing
more compounds are more likely to be used in specific cuisines. The results
largely correlate with our earlier observations: in North American recipes, the
more compounds are shared by two ingredients, the more likely they appear in
recipes. By contrast, in East Asian cuisine the more flavor compounds two
ingredients share, the less likely they are used together (Fig.~3G and~3H; see
SI for details and results on other cuisines).

What is the mechanism responsible for these differences? That is, does Fig.~3C
through H imply that all recipes aim to pair ingredients together that share
(North America) or do not share (East Asia) flavor compounds, or could we
identify some compounds responsible for the bulk of the observed effect? We
therefore measured the contribution $\chi_i$ of each ingredient to the shared
compound effect in a given cuisine $c$, quantifying to what degree its presence
affects the magnitude of $\Delta N_s $. 

In Fig.~3I,J we show as a scatter plot $\chi_i$ (horizontal axis) and the
frequency $f_i$ for each ingredient in North American and East Asian cuisines.
The vast majority of the ingredients lie on the $\chi_i=0$ axis, indicating
that their contribution to $\Delta N_{s}$ is negligible. Yet, we observe a few
frequently used outliers, which tend to be in the positive $\chi_i$ region for
North American cuisine, and lie predominantly in the negative region for East
Asian cuisine.  This suggests that the food pairing effect is due to a few
outliers that are frequently used in a particular cuisine, e.g. milk, butter,
cocoa, vanilla, cream, and egg in the North America, and beef, ginger, pork,
cayenne, chicken, and onion in East Asia.  Support for the definitive role of
these ingredients is provided in Fig.~3K,L where we removed the ingredients in
order of their positive (or negative) contributions to $\Delta N_{s}$ in the
North American (or East Asian) cuisine, finding that the $z$-score, which
measures the significance of the shared compound hypothesis, drops below two
after the removal of only 13 (5) ingredients from North American (or East
Asian) cuisine (see SI S2.2.2).  Note, however, that these ingredients play a
disproportionate role in the cuisine under consideration---for example, the 13
key ingredients contributing to the shared compound effect in North American
cuisine appear in
74.4\% of all recipes.

According to an empirical view known as ``the flavor
principle''~\cite{flavor_principle}, the differences between regional cuisines
can be reduced to a few key ingredients with specific flavors: adding soy sauce
to a dish almost automatically gives it an oriental taste because Asians use
soy sauce widely in their food and other ethnic groups do not; by contrast
paprika, onion, and lard is a signature of Hungarian cuisine. Can we
systematically identify the ingredient combinations responsible for the taste
palette of a regional cuisine? To answer this question, we measure the
\emph{authenticity} of each ingredient ($p^c_{i}$), ingredient pair
($p^c_{ij}$), and ingredient triplet ($p^c_{ijk}$) (see Materials and Methods).
In Fig.~4 we organize the six most authentic single ingredients, ingredient
pairs and triplets for North American and East Asian cuisines in a flavor
pyramid. The rather different ingredient classes (as reflected by their color)
in the two pyramids capture the differences between the two cuisines: North
American food heavily relies on dairy products, eggs and wheat; by contrast,
East Asian cuisine is dominated by plant derivatives like soy sauce, sesame
oil, and rice and ginger.  Finally, the two pyramids also illustrate the
different affinities of the two regional cuisines towards food pairs with
shared compounds. The most authentic ingredient pairs and triplets in the North
American cuisine share multiple flavor compounds, indicated by black links, but
such compound-sharing links are rare among the most authentic combinations in
East Asian cuisine. 

The reliance of regional cuisines on a few authentic ingredient combinations
allows us to explore the ingredient-based relationship (similarity or
dissimilarity) between various regional cuisines. For this we selected the six
most authentic ingredients and ingredient pairs in each regional cuisine (i.e.
those shown in Fig.~4A,B), generating a diagram that illustrates the
ingredients shared by various  cuisines, as well as singling out those that are
unique to a particular region (Fig.~4C). We once again find a close
relationship between North American and Western European cuisines and observe
that when it comes to its signature ingredient combinations Southern European
cuisine is much closer to Latin American than Western European cuisine
(Fig.~4C). 

\section*{Discussion}

Our work highlights the limitations of the recipe data sets currently
available, and more generally of the systematic analysis of food preparation
data. By comparing two editions of the same dataset with significantly
different coverage, we can show that our results are robust against data
incompleteness (see SI S1.1.2). Yet, better compound databases, mitigating the
incompleteness and the potential biases of the current data, could
significantly improve our understanding of food. There is inherent ambiguity in
the definition of a particular regional or ethnic cuisine. However, as
discussed in SI S1.2, the correlation between different datasets, representing
two distinct perspectives on food (American and Korean), indicates that humans
with different ethnic background have a rather consistent view on the
composition of various regional cuisines.

Recent work by Kinouchi et al.~\cite{kinouchi_culinaryevol_2008} observed that
the frequency-rank plots of ingredients are invariant across four different
cuisines, exhibiting a shape that can be well described by a Zipf-Mandelbrot
curve. Based on this observation, they model the evolution of recipes by
assuming a copy-mutate process, leading to a very similar frequency-rank curve.
The copy-mutate model provides an explanation for how an ingredient becomes a
staple ingredient of a cuisine: namely, having a high \emph{fitness} value or
being a founder. The model assigns each ingredient a random fitness value,
which represents the ingredient's nutritional value, availability, flavor, etc.
For example, it has been suggested that each culture eagerly adopt spices that
have high anti-bacterial activity (e.g.  garlic)~\cite{billing1998,
sherman2001}, spices considered to have high fitness. The mutation phase of the
model replaces less fit ingredients with more fit ones.  Meanwhile, the copy
mechanism keeps copying the \emph{founder ingredients}---ingredients in the
early recipes---and makes them abundant in the recipes regardless of their
fitness value.

It is worthwhile to discuss the similarity and difference between the
quantities we measured and the concepts of fitness and founders.  First of all,
prevalence ($P^c_i$) and authenticity ($p^c_i$) are empirically measured values
while fitness is an intrinsic hidden variable. Among the list of highly
prevalent ingredients we indeed find \emph{old} ingredients---founders---that
have been used in the same geographic region for thousands of years.  At the
same time, there are relatively new ingredients such as tomatoes, potatoes, and
peppers that were introduced to Europe and Asia just a few hundred years ago.
These new, but prevalent ingredients can be considered to have high fitness
values. If an ingredient has a high level of authenticity, then it is prevalent
in a cuisine while not so prevalent in all other cuisines. 

Indeed, each culture has developed their own authentic ingredients. It may
indicate that fitness can vary greatly across cuisines or that the
stochasticity of recipe evolution diverge the recipes in different regions into
completely different sets. More historical investigation will help us to
estimate the fitness of ingredients and assess why we use the particular
ingredients we currently do. The higher order fitness value suggested
in~\cite{kinouchi_culinaryevol_2008} is very close to our concept of food
pairing affinity.

Another difference in our results is the number of ingredients in recipes.
Kinouchi et al. reported that the average number of ingredients per recipe
varies across different cookbooks.  While we also observed variation in the
number of ingredients per recipe, the patterns we found were not consistent
with those found by Kinouchi et al.  For instance, the French cookbook has more
ingredients per recipe than a Brazillian one, but in our dataset we find the
opposite result. We believe that a cookbook cannot represent a whole cuisine,
and that cookbooks with more sophisticated recipes will tend to have more
ingredients per recipe than cookbooks with everyday recipes. As more complete
datasets become available, sharper conclusions can be drawn regarding the size
variation between cuisines.  

Our contribution in this context is a study of the role that flavour compounds
play in determining these fitness values. One possible interpretation of our
results is that shared flavor compounds represent one of several contributions
to fitness value, and that, while shared compounds clearly play a significant
role in some cuisines, other contributions may play a more dominant role in
other cuisines. The fact that recipes rely on ingredients not only for flavor
but also to provide the final textures and overall structure of a given dish
provides support for the idea that fitness values depend on a multitude of
ingredient characteristics besides their flavor profile.

In summary, our network-based investigation identifies a series of
statistically significant patterns that characterize the way humans choose the
ingredients they combine in their food. These patterns manifest themselves to
varying degree in different geographic regions: while North American and
Western European dishes tend to combine ingredients that share flavor
compounds, East Asian cuisine avoids them. More generally this work provides an
example of how the data-driven network analysis methods that have transformed
biology and the social sciences in recent years can yield new insights in other
areas, such as food science.

\section*{Methods}

\subsection*{Shared compounds}

To test the hypothesis that the choice of ingredients is driven by an
appreciation for ingredient pairs that share flavor compounds (i.e. those
linked in Fig.~2), we measured the mean number of shared compounds in each
recipe, $N_s$, comparing it with $N^\mathrm{rand}_s$ obtained for a randomly
constructed reference recipe dataset. For a recipe $R$ that contains $n_R$
different ingredients, where each ingredient $i$ has a set of flavor compounds
$C_i$, the mean number of shared compounds 
\begin{equation}
N_s(R) = \frac{2}{n_R(n_R-1)} \sum_{i,j \in R, i \neq j} \left | C_i \cap C_j
\right | \label{eq:ns}
\end{equation}
is zero if none of the ingredient pairs $(i, j)$ in the recipe share any flavor
compounds. For example, the `mustard cream pan sauce' recipe contains chicken
broth, mustard, and cream, none of which share any flavor compounds ($N_s (R) =
0$) in our dataset. Yet, $N_s (R)$ can reach as high as 60 for `sweet and
simple pork chops', a recipe containing apple, pork, and cheddar cheese (See
Fig.~3A). To check whether recipes with high $N_s(R)$ are statistically
preferred (implying the validity of the shared compound hypothesis) in a
cuisine $c$ with $N_c$ recipes, we calculate $\Delta N_s = N^\mathrm{real}_s -
N^\mathrm{rand}_s$, where `real' and `rand' indicates real recipes and randomly
constructed recipes respectively and $N_s = \sum_{R}N_s(R) / N_c$ (see SI for
details of the randomization process). This random reference (null model)
controls for the frequency of a particular ingredient in a given regional
cuisine, hence our results are not affected by historical, geographical, and
climate factors that determine ingredient availability (see SI S1.1.2).

\subsection*{Contribution} The contribution $\chi_i$ of each ingredient to the
shared compound effect in a given cuisine $c$, quantifying to what degree its
presence affects the magnitude of $\Delta N_s $, is defined by 
%
\begin{equation} \chi_{i} = \left ( \frac{1}{N_c} \sum_{R \ni i} \frac{2}{n_R
(n_R - 1)} \sum_{j \neq i (j,i \in R)} \left | C_{i} \cap C_{j} \right |
\right) - \left ( \frac{2 f_i}{N_c \langle n_R \rangle} \frac{\sum_{j \in c}
f_j \left | C_{i} \cap C_{j} \right |}{\sum_{j \in c} f_j} \right ),
\label{eq:contribution} \end{equation}
where $f_i$ represents the ingredient $i$'s number of occurrence.  An
ingredient's contribution is positive (negative) if it increases (decreases)
$\Delta N_s$. 

\subsection*{Authenticity} we define the prevalence $P^c_i$ of each
ingredient $i$ in a cuisine $c$ as $P^c_i = n^c_i / N_c$, where $n^c_i$ is the
number of recipes that contain the particular ingredient $i$ in the cuisine and
$N_c$ is the total number of recipes in the cuisine. The relative
prevalence $p^c_{i} = P^c_{i} - \langle P^{c'}_i \rangle_{c'\neq c}$ measures
the authenticity---the difference between the prevalence of $i$ in
cuisine $c$ and the average prevalence of $i$ in all other cuisines.  We can
also identify ingredient pairs or triplets that are overrepresented in a
particular cuisine relative to other cuisines by defining the relative pair
prevalences $p^c_{ij} = P^c_{ij} - \langle P^{c'}_{ij}\rangle_{c'\neq c}$ and
triplet prevalences $p^c_{ijk} = P^c_{ijk} - \langle P^{c'}_{ijk}
\rangle_{c'\neq c}$, with $P^c_{ij} = n^c_{ij} / N_c$ and $P^c_{ijk} =
n^c_{ijk} / N_c$.  

\bibliographystyle{naturemag}

\section*{Acknowledgements}

We thank M.~I. Meirelles, S. Lehmann, D. Kudayarova, T.~W. Lim, J. Baranyi, H.
This for discussions. This work was supported by the James S. McDonnell
Foundation 21st Century Initiative in Studying Complex Systems. 

\section*{Author contributions}

YYA, SEA, and ALB designed research and YYA, SEA, and JPB performed research. 
All authors wrote and reviewed the manuscript.

\clearpage

\begin{figure}\centering
    \includegraphics[width=\textwidth]{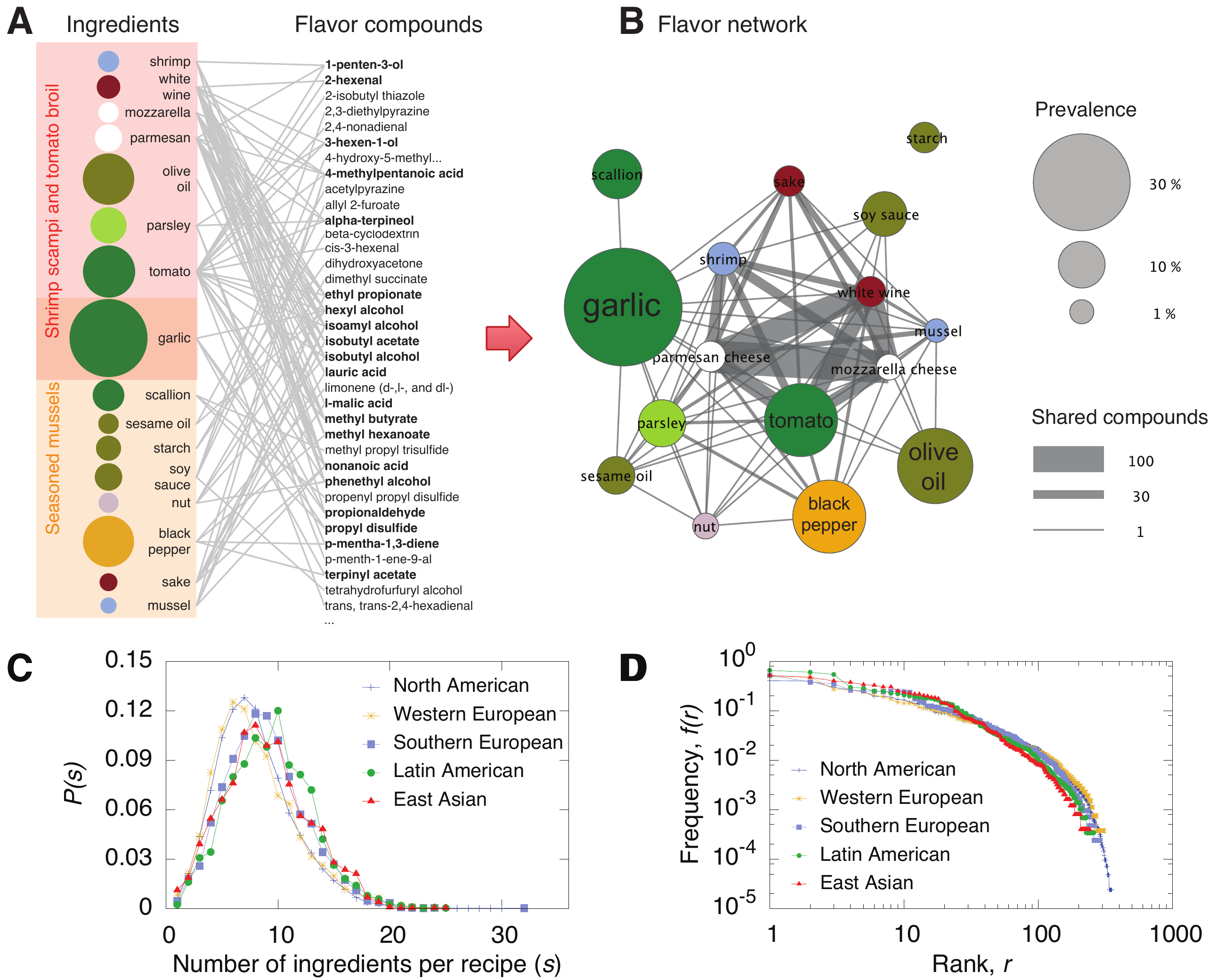}

    \caption[]{\textbf{Flavor network.} \letter{A} The ingredients contained in
two recipes (left column), together with the flavor compounds that are known to
be present in the ingredients (right column). Each flavor compound is linked to
the ingredients that contain it, forming a bipartite network.  Some compounds
(shown in boldface) are shared by multiple ingredients.  \letter{B} If we
project the ingredient-compound bipartite network into the ingredient space, we
obtain the \emph{flavor network}, whose nodes are ingredients, linked if they
share at least one flavor compound. The thickness of links represents the
number of flavor compounds two ingredients share and the size of each circle
corresponds to the prevalence of the ingredients in recipes. \letter{C} The
distribution of recipe size, capturing the number of ingredients per recipe,
across the five cuisines explored in our study.  \letter{D} The frequency-rank
plot of ingredients across the five cuisines show an approximately invariant
distribution across cuisines. \label{fig:recipe_ingredient_compound}}

\end{figure}

\begin{landscape}
\begin{figure}\centering

    \includegraphics[width=\columnwidth, trim=0 0 0 0, clip=true]{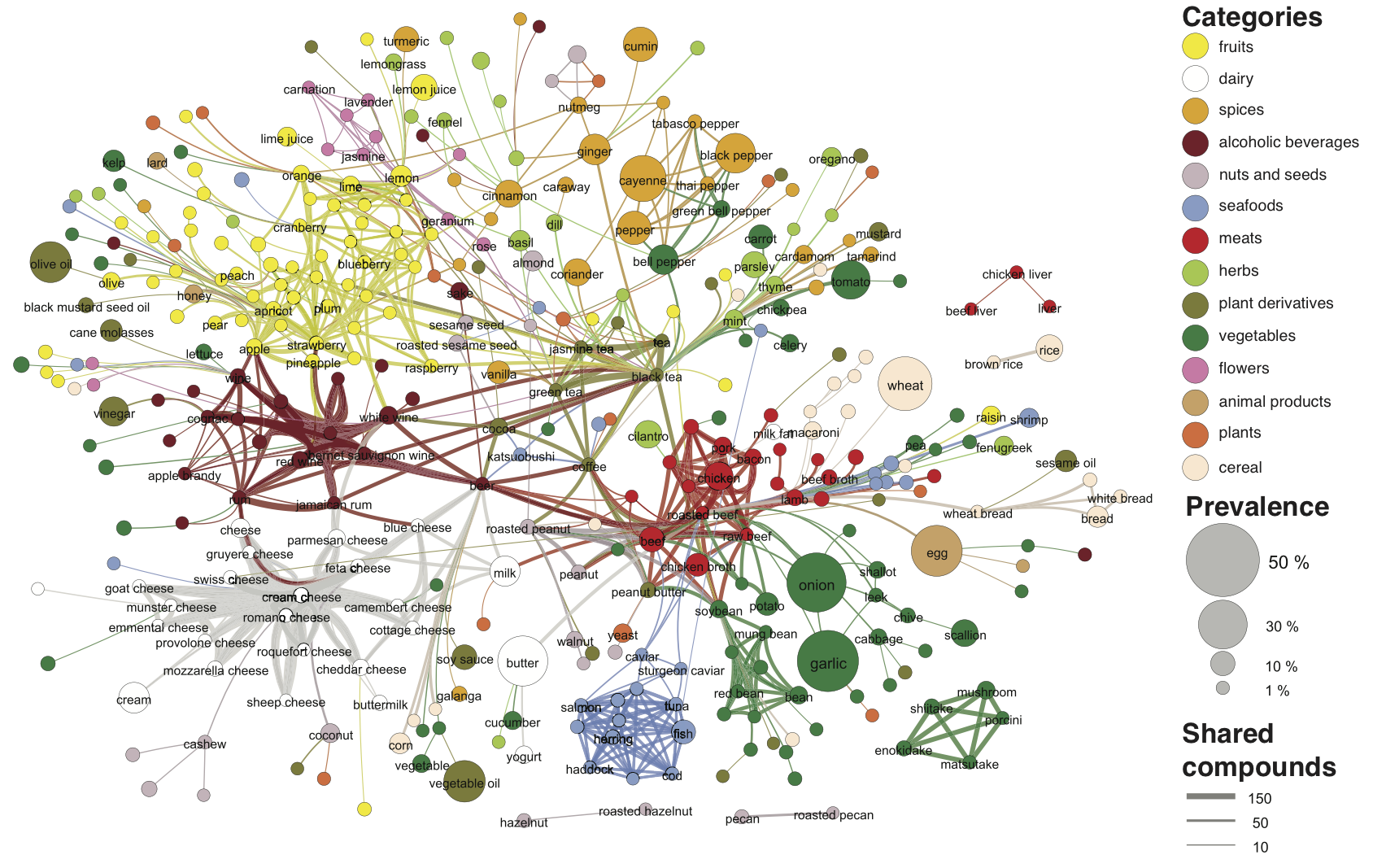}

    \caption[]{\textbf{The backbone of the flavor network.} Each node denotes an
    ingredient, the node color indicates food category, and node size reflects
    the ingredient prevalence in recipes. Two ingredients are connected if they
    share a significant number of flavor compounds, link thickness representing
    the number of shared compounds between the two ingredients. Adjacent links
    are bundled to reduce the clutter. Note that the map shows only the
    statistically significant links, as identified by the algorithm of
    Refs.\cite{serrano2009, shlee_google}  for $p$-value 0.04. A drawing of the
    full network is too dense to be informative. We use, however, the full
    network in our subsequent measurements. 
    \label{fig:network1}} 

\end{figure}
\end{landscape}

\begin{landscape}
\begin{figure}\centering
    \includegraphics[width=0.7\columnwidth, trim=0 0 0 0, clip=true]{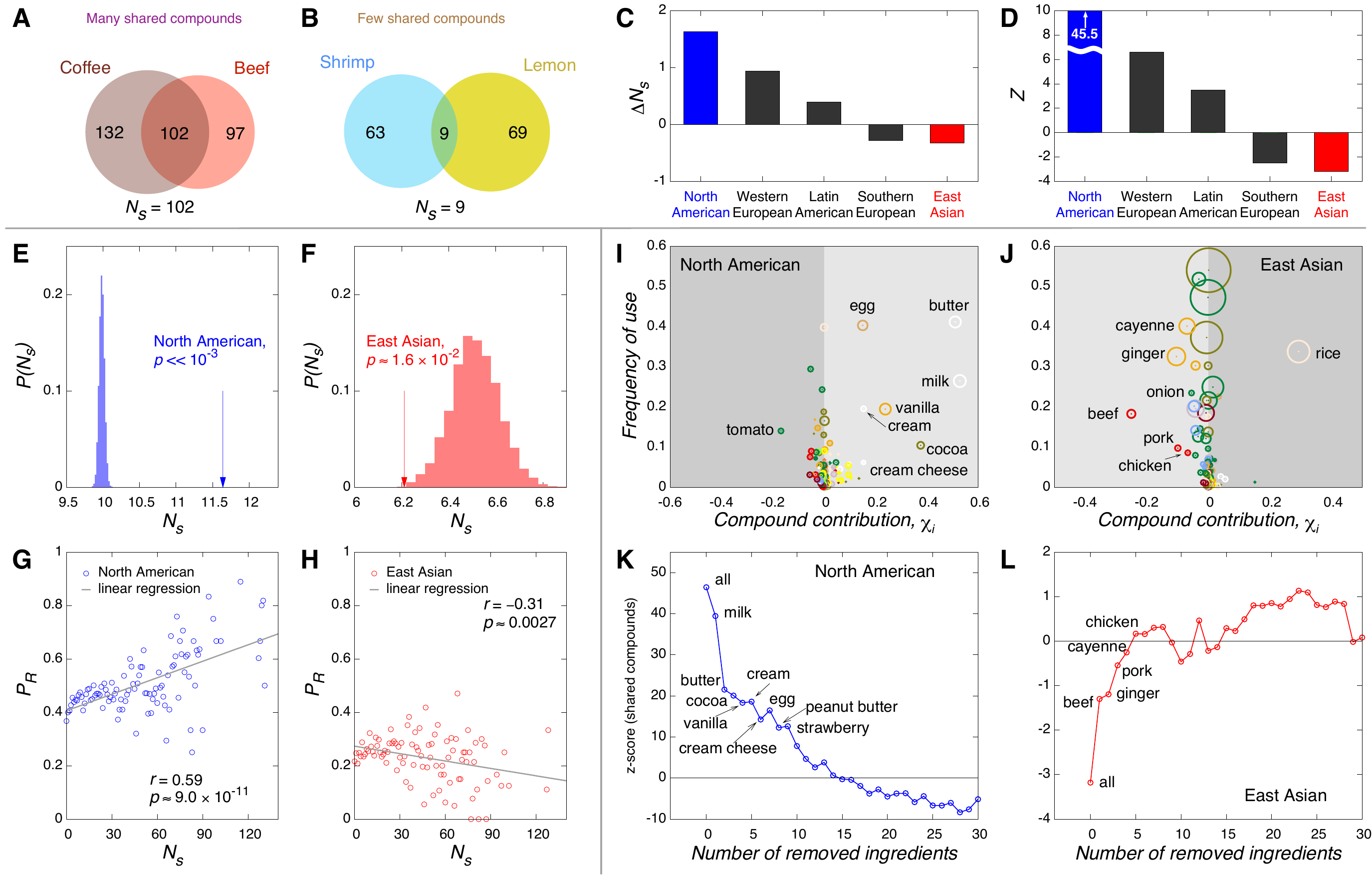}

    \caption[]{\textbf{Testing the food pairing hypothesis.} Schematic
    illustration of two ingredient pairs, the first sharing many more
    \letter{A} and the second much fewer \letter{B} compounds than expected if
    the flavor compounds were distributed randomly. \letter{C,D} To test the
    validity of the food pairing hypothesis, we construct 10,000 random recipes
    and calculate $\Delta N_s$. 
    We find that ingredient pairs in North American cuisines tend to share more
    compounds while East Asian cuisines tend to share fewer compounds than
    expected in a random recipe dataset. \letter{E,F} The distributions
    $P(N_s)$  for 10,000 randomized recipe datasets compared with the real
    values for East Asian and North American cuisine. Both cuisines exhibit
    significant $p$-values, as estimated using a $z$-test. \letter{G,H} We
    enumerate every possible ingredient pair in each cuisine and show the
    fraction of pairs in recipes as a function of the number of shared
    compounds. To reduce noise, we only used data points calculated from more
    than 5 pairs. The $p$-values are calculated using a $t$-test. North
    American cuisine is biased towards pairs with more shared compounds while
    East Asian shows the opposite trend (see SI for details and results for
    other cuisines). Note that we used the full network, not the backbone shown
    in Fig.~\ref{fig:network1} to obtain these results.  \letter{I,J} The
    contribution and frequency of use for each ingredient in North American and
    East Asian cuisine.  The size of the circles represents the relative
    prevalence $p_i^c$. North American and East Asian cuisine shows the
    opposite trends.  \letter{K,L} If we remove the highly contributing
    ingredients sequentially (from the largest contribution in North American
    cuisine and from the smallest contribution in East Asian cuisine), the
    shared compounds effect quickly vanishes when we removed five (East Asian)
    to fifteen (North American) ingredients.  \label{fig:hypothesis}}

\end{figure}
\end{landscape}

\begin{figure}
    \begin{minipage}{0.5\textwidth}
    \includegraphics[width=\textwidth, trim=0 0 0 0, clip=true]{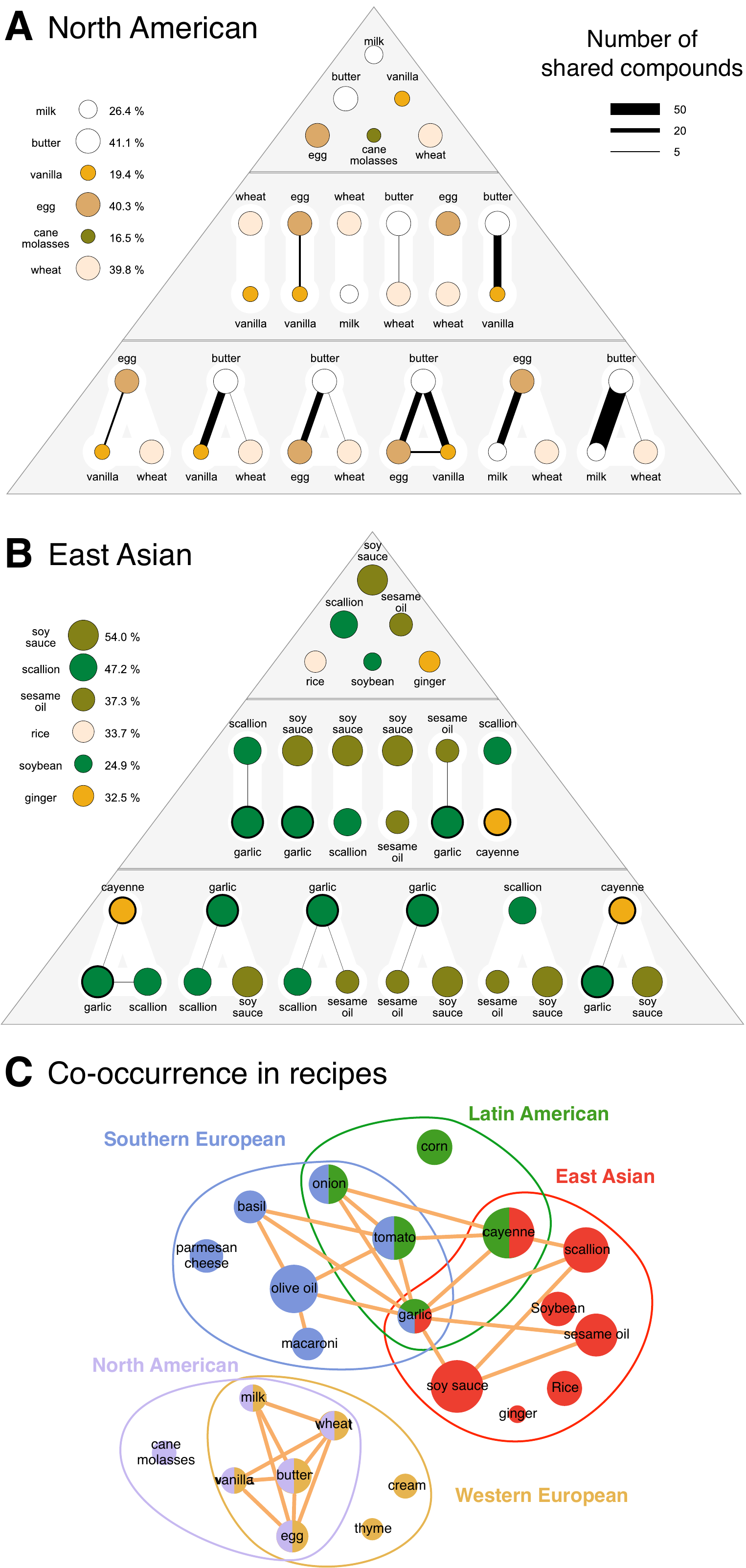}
    \end{minipage}

    \caption[]{\textbf{Flavor principles.} \letter{A,B} Flavor pyramids for North
    American and East Asian cuisines. Each flavor pyramid shows the six most
    \textbf{authentic} ingredients (i.e. those with the largest $p^c_{i}$),
    ingredient pairs (largest $p^c_{ij}$), and ingredient triplets (largest
    $p^c_{ijk}$). The size of the nodes reflects the abundance $P^c_{i}$ of the
    ingredient in the recipes of the particular cuisine. Each color represents
    the category of the ingredient (see Fig.~\ref{fig:network1} for the color)
    and link thickness indicates the number of shared compounds. \letter{C} The
    six most authentic ingredients and ingredient pairs used in specific
    regional cuisine. Node color represents cuisine and the link weight
    reflects the relative prevalence $p^c_{i}$ of the ingredient pair.
    \label{fig:flavor_pyramid}} 
    
\end{figure} 

\clearpage


\setcounter{page}{1}
\setcounter{figure}{0}
\setlength{\parskip}{9pt}

\renewcommand{\thefigure}{S\arabic{figure}}
\renewcommand{\thesection}{S\arabic{section}}
\renewcommand{\thetable}{S\arabic{table}}
\renewcommand{\theequation}{S\arabic{equation}}

\singlespacing

{\huge \noindent Supporting Online Material}\\
\baselineskip 24pt
\noindent \emph{Flavor network and the principles of food pairing} \\
\noindent by Yong-Yeol Ahn, Sebastian E.~Ahnert, James P.~Bagrow, Albert-L\'aszl\'o Barab\'asi

\renewcommand*\contentsname{Table of Contents}
\begin{spacing}{0.9}
\tableofcontents
\listoffigures
\listoftables
\end{spacing}

\section{Materials and methods} 

\subsection{Flavor network} 

\subsubsection{Ingredient-compounds bipartite network} 

The starting point of our research is Fenaroli's handbook of flavor ingredients
(fifth edition~\cite{Sfenaroli}), which offers a systematic list of flavor
compounds and their natural occurrences (food ingredients). Two post-processing
steps were necessary to make the dataset appropriate for our research: (A) In
many cases, the book lists the essential oil or extract instead of the
ingredient itself. Since these are physically extracted from the original
ingredient, we associated the flavor compounds in the oils and extracts with
the original ingredient. (B) Another post-processing step is including the
flavor compounds of a more general ingredient into a more specific ingredient.
For instance, the flavor compounds in `meat' can be safely assumed to also be
in `beef' or `pork'. `Roasted beef' contains all flavor compounds of `beef' and
`meat'.

The ingredient-compound association extracted from~\cite{Sfenaroli} forms a
bipartite network. As the name suggests, a bipartite network consists of two
types of nodes, with connections only between nodes of different types.  Well
known examples of bipartite networks include collaboration networks of
scientists \cite{Snewman01} (with scientists and publications as nodes) and
actors \cite{Swattsstrogatz98} (with actors and films as nodes), or the human
disease network \cite{Skwang07} which connects health disorders and disease
genes. In the particular bipartite network we study here, the two types of
nodes are food ingredients and flavor compounds, and a connection signifies
that an ingredient contains a compound. 

\begin{figure}[b!] 

    \begin{minipage}[t]{0.5\textwidth}
	\centering
	\includegraphics[width=1\textwidth]{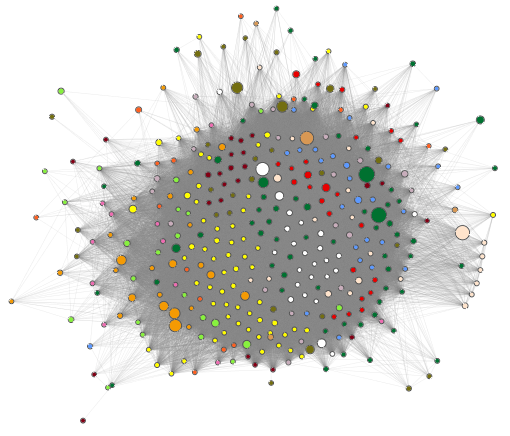}
    \end{minipage}
    \hspace{0.02\textwidth}
    \begin{minipage}[b]{0.45\textwidth}

	\caption[Full ingredient network]{The full flavor network. The size of
	a node indicates average prevalence, and the thickness of a link
	represents the number of shared compounds. All edges are drawn. It is
	impossible to observe individual connections or any modular structure.
	\label{fig:food_network}} \vspace{1.9cm}

    \end{minipage}

\end{figure} 
\begin{figure}[b!]\centering 
    \includegraphics[width=0.8\textwidth]{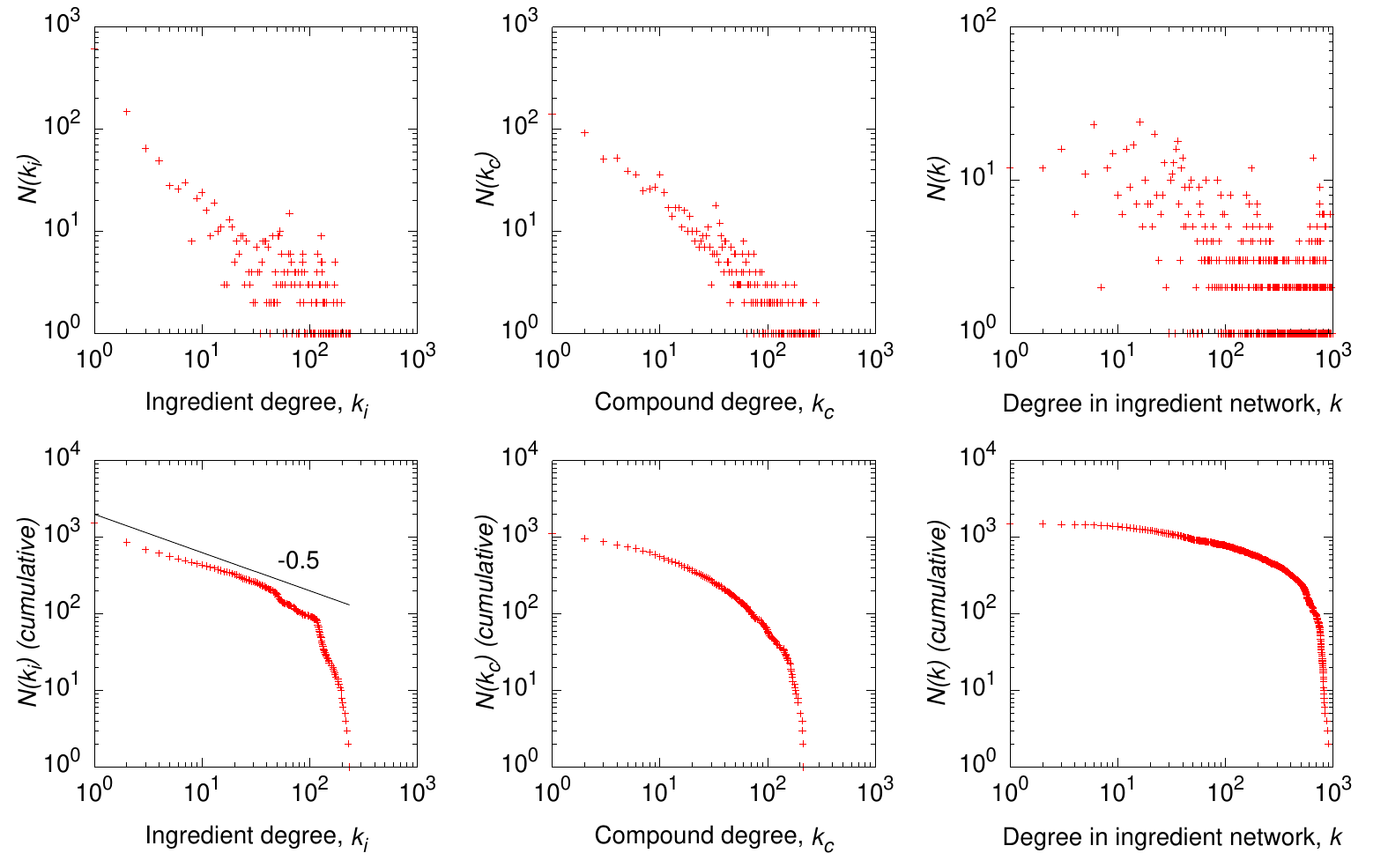}

    \caption[Degree distribution of flavor network]{Degree distributions of the
flavor network. Degree distribution of ingredients in the ingredient-compound
network, degree distribution of flavor compounds in the ingredient-compound
network, and degree distribution of the (projected) ingredient network, from
left to right. Top: degree distribution. Bottom: complementary cumulative
distribution. The line and the exponent value in the leftmost figure at the
bottom is purely for visual guide.\label{fig:degree_dist}}

\end{figure} 

The full network contains 1,107 chemical compounds and 1,531 ingredients, but
only 381 ingredients appear in recipes, together containing 1,021 compounds
(see Fig.~\ref{fig:food_network}). We project this network into a weighted
network between ingredients only~\cite{Snewman_structure_2006,
Scaldarelli_scale-free_2007, Sdoro_book,barrat_weighted}. The weight of each edge
$w_{ij}$ is the number of compounds shared between the two nodes (ingredients)
$i$ and $j$, so that the relationship between the $M \times M$ weighted
adjacency matrix $w_{ij}$ and the $N \times M$ bipartite adjacency matrix
$a_{ik}$ (for ingredient $i$ and compound $k$) is given by: \begin{equation}
w_{ij} = \sum_{k=1}^N a_{ik} a_{jk} \end{equation}

The degree distributions of ingredients and compounds are shown in
Fig.~\ref{fig:degree_dist}.


\subsubsection{Incompleteness of data and the third edition}\label{sec:incompleteness}  

\begin{figure}\centering 

    \includegraphics[width=0.9\textwidth]{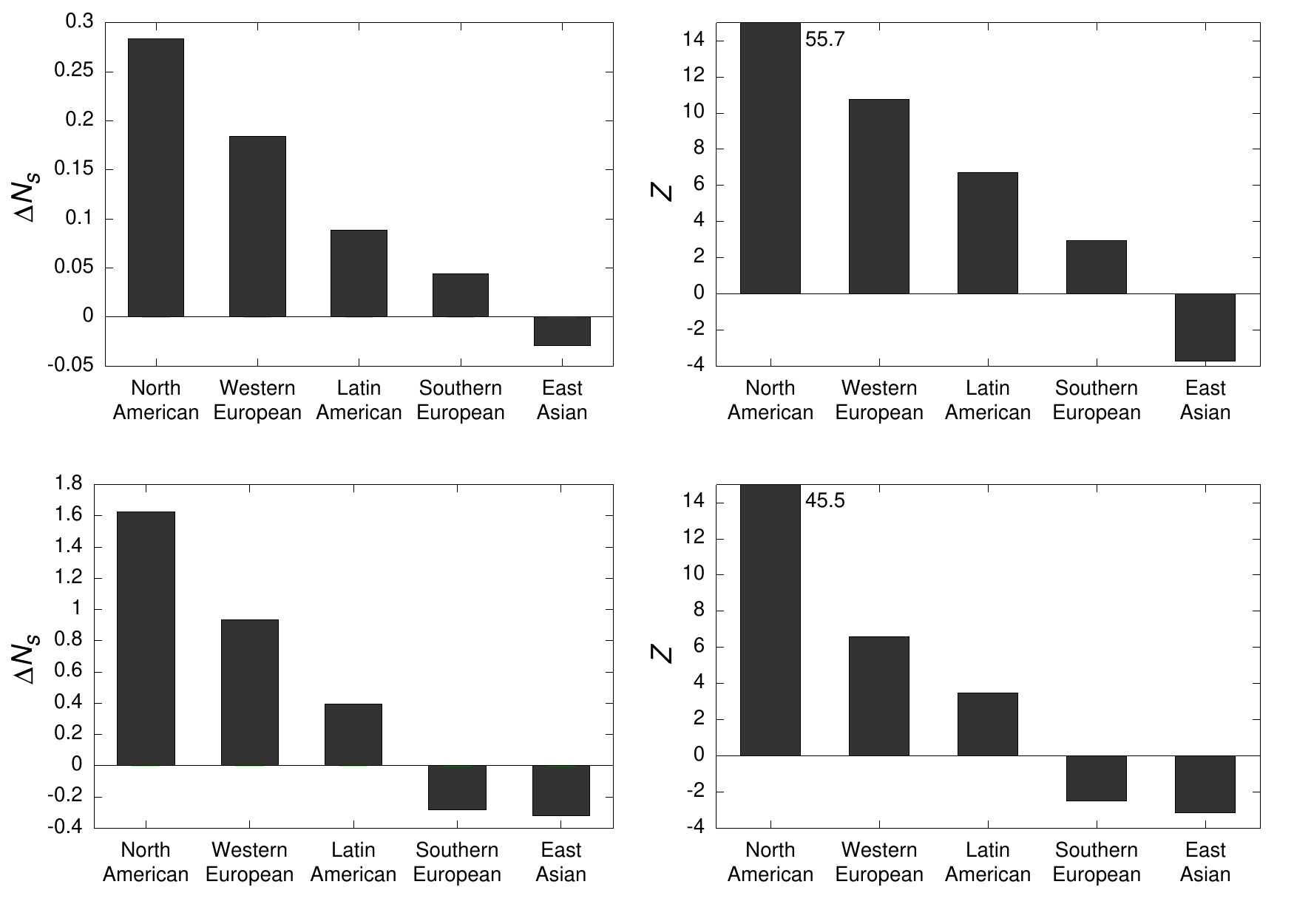}

    \caption[Comparing the third and fifth edition of Fenaroli's] {Comparing
    the third and fifth edition of Fenaroli's to see if incomplete data impacts
    our conclusions. The much sparser data of the 3rd edition (\textbf{Top})
    shows a very similar trend to that of the 5th edition (\textbf{Bottom},
    repeated from main text Fig.~3). Given the huge difference between the two
    editions (Table~\ref{tab:third}), this further supports that the observed
    patterns are robust. \label{fig:third}}

\end{figure} 
\begin{table}\centering 
\begin{tabular}{c|c|c}

                           & 3rd eds. & 5th eds. \\\hline
\# of ingredients          & 916      & 1507 \\
\# of compounds            & 629      & 1107 \\
\# of edges in I-C network & 6672     & 36781 \\

\end{tabular}

\caption[Statistics of 3rd and 5th editions]{The basic statistics on two
different datasets. The 5th Edition of Fenaroli's handbook contains much more
information than the third edition.\label{tab:third}}

\end{table} 

The situation encountered here is similar to the one encountered in systems
biology: we do not have a complete database of all protein, regulatory and
metabolic interactions that are present in the cell. In fact, the existing
protein interaction data covers \emph{less than 10\% of all protein
interactions} estimated to be present in the human cell~\cite{Svenkatesan2009}.

To test the robustness of our results against the incompleteness of data, we
have performed the same calculations for the 3rd edition of Fenaroli's handbook
as well. The 5th edition contains approximately six times more information on
the chemical contents of ingredients (Table~\ref{tab:third}). Yet, our main
result is robust (Fig.~\ref{fig:third}), further supporting that data
incompleteness is not the main factor behind our findings.


\subsubsection{Extracting the backbone} 

\begin{figure}[t!]\centering 

    \framebox{\includegraphics[width=0.7\textwidth, trim=0 0 0 0,
    clip=true]{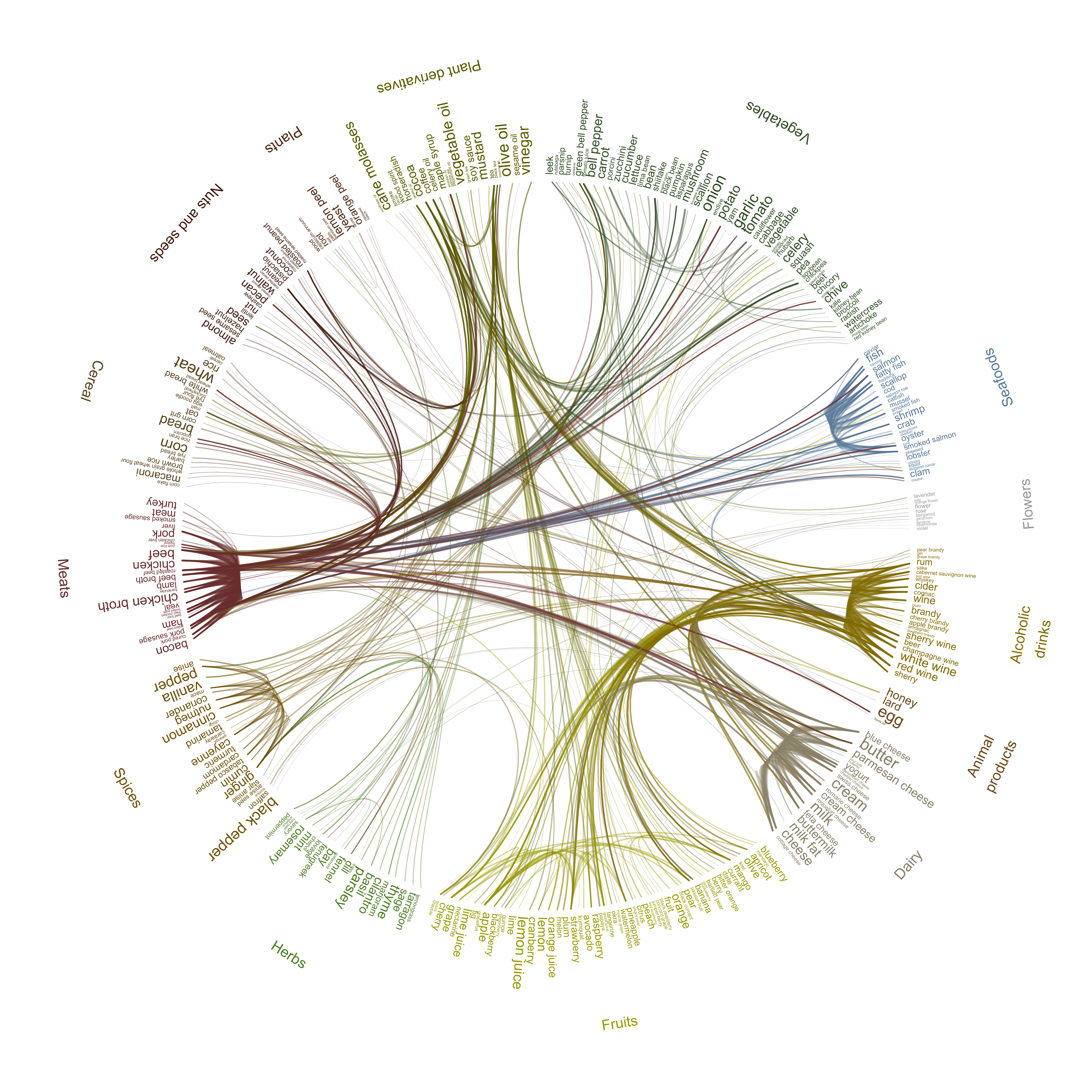}}

    \caption[Backbone]{The backbone of the ingredient network extracted
    according to~\cite{Sserrano2009} with a significance threshold $p = 0.04$.
    Color indicates food category, font size reflects ingredient prevalence in
    the dataset, and link thickness represents the number of shared compounds
    between two ingredients. 
    \label{fig:backbone}}

\end{figure} 

The network's average degree is about 214 (while the number of nodes is 381).
It is very dense and thus hard to visualize (see Fig.~\ref{fig:food_network}).
To circumvent this high density, we use a method that extracts the backbone of
a weighted network~\cite{Sserrano2009}, along with the method suggested
in~\cite{Sshlee_google}. For each node, we keep those edges whose weight is
statistically significant given the strength (sum of weight) of the node. If
there is none, we keep the edge with the largest weight. A different
visualization of this backbone is presented in Fig.~\ref{fig:backbone}.
Ingredients are grouped into categories and the size of the name indicates the
prevalence. This representation clearly shows the categories that are closely
connected.

\subsubsection{Sociological bias} 

\begin{figure}\centering 

    \includegraphics[width=\textwidth]{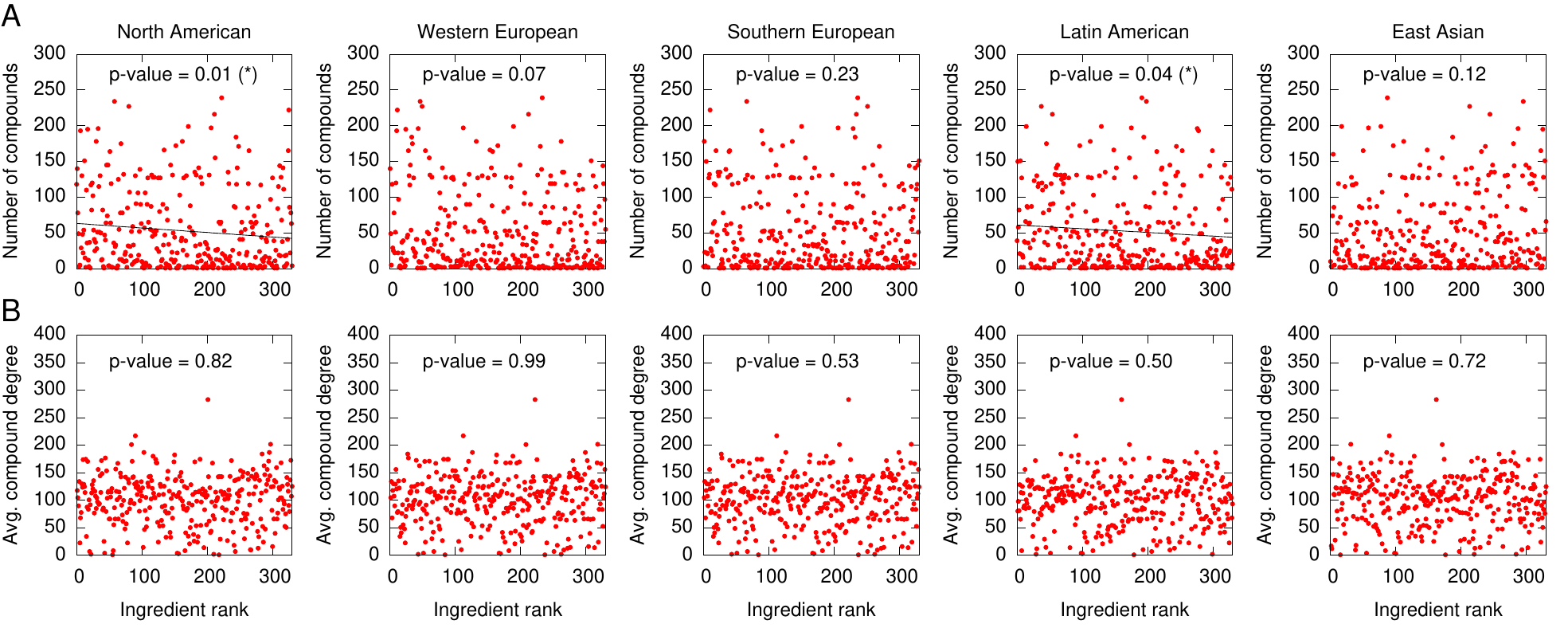}

    \caption[Potential biases]{Are popular, much-used ingredients more studied
    than less frequent foods, leading to potential systematic bias? \letter{A}
    We plot the number of flavor compounds for each ingredient as a function of
    the (ranked) popularity of the ingredient. The correlation is very small
    compared to the large fluctuations present. There is a weak tendency that
    the ingredients mainly used in North American or Latin American cuisine
    tend to have more odorants, but the correlations are weak (with
    coefficients of -0.13 and -0.10 respectively).  A linear regression line is
    shown only if the corresponding $p$-value is smaller than $0.05$.
    \letter{B} If there is bias such that the book tends to list more familiar
    ingredients for more common flavor compounds, then we can observe the
    correlation between the familiarity (how frequently it is used in the
    cuisine) and the degree of the compound in the ingredient-compound network.
    The plots show no observable correlations for any cuisine.
    \label{fig:possible_bias}}

\end{figure}

Western scientists have been leading food chemistry, which may imply that
western ingredients are more studied. To check if such a bias is present in our
dataset, we first made two lists of ingredients: one is the list of ingredients
appearing in North American cuisine, sorted by the relative prevalence $p^c_i$
(i.e. the ingredients more specific to North American cuisine comes first). The
other is a similar list for East Asian cuisine. Then we measured the number of
flavor compounds for ingredients in each list. The result in
Fig.~\ref{fig:possible_bias}A shows that any potential bias, if present, is not
significant. 

There is another possibility, however, if there is bias such that the dataset 
tends to list more familiar (Western) ingredients for more common flavor
compounds, then we should observe a correlation between the familiarity
(frequently used in Western cuisine) and the degree of compound (number of
ingredients it appears in) in the ingredient. Figure~\ref{fig:possible_bias}B
shows no observable correlation, however.

\subsection{Recipes}\label{sec:recipes} 

The number of potential ingredient combinations is enormous. For instance, one
could generate $\sim 10^{15}$ distinct ingredient combinations by choosing
eight ingredients (the current average per recipe) from approximately 300
ingredients in our dataset.  If we use the numbers reported in Kinouchi et
al.~\cite{Skinouchi_culinaryevol_2008} (1000 ingredients and 10 ingredients per
recipe), one can generate $\sim 10^{23}$ ingredient combinations. This number
greatly increases if we consider the various cooking methods. Regardless, the
fact that this number exceeds by many orders of magnitude the
$\sim 10^6$ recipes listed in the largest recipe repositories (e.g.
\url{http://cookpad.com}) indicates that humans are exploiting a tiny fraction
of the culinary space. 

\begin{figure}\centering 

    \includegraphics[width=0.6\textwidth]{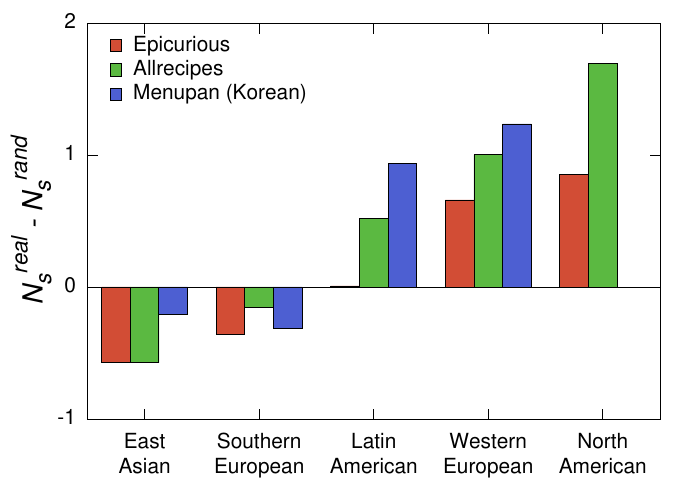}

    \caption[Coherency of datasets]{Comparison between different datasets. The
    results on different datasets qualitatively agree with each other (except
    Latin American cuisine). Note that \emph{menupan.com} is a Korean
    website.  \label{fig:coherency}}

\end{figure}

We downloaded all available recipes from three websites: \emph{allrecipes.com},
\emph{epicurious.com}, and \emph{menupan.com}.  Recipes tagged as belonging to
an ethnic cuisine are extracted and then grouped into 11 larger regional
groups. We used only 5 groups that each contain more than 1,000 recipes (See
Table~\ref{tab:cuisine_stat}). In the curation process, we made a replacement
dictionary for frequently used phrases that should be discarded, synonyms for
ingredients, complex ingredients that are broken into ingredients, and so
forth. We used this dictionary to automatically extract the list of ingredients
for each recipe. As shown in Fig.~1D, the usage of ingredients is highly
heterogenous. Egg, wheat, butter, onion, garlic, milk, vegetable oil, and cream
appear more than 10,000 recipes while geranium, roasted hazelnut, durian,
muscat grape, roasted pecan, roasted nut, mate, jasmine tea, jamaican rum,
angelica, sturgeon caviar, beech, lilac flower, strawberry jam, and emmental
cheese appear in only one recipe.  Table~\ref{tab:corr_data} shows the
correlation between ingredient usage frequency in each cuisine and in each
dataset.  Figure.~\ref{fig:coherency} shows that the three datasets
qualitatively agree with each other, offering a base to combine these datasets.

\begin{table} 
    \centering

    \caption[Recipe dataset]{Number of recipes and the detailed cuisines in
    each regional cuisine in the recipe dataset. Five groups have reasonably
    large size. We use all cuisine data when calculating the relative
    prevalence and flavor principles. }

    \begin{tabular}{r | r | p{8cm} }
	\hline\hline
	Cuisine set  &   Number of recipes & Cuisines included \\
	\hline
	North American    & 41525 & American, Canada, Cajun, Creole, Southern soul food, Southwestern U.S.    \\
	Southern European & 4180  & Greek, Italian, Mediterranean, Spanish, Portuguese                        \\
	Latin American    & 2917  & Caribbean, Central American, South American, Mexican                      \\
	Western European  & 2659  & French, Austrian, Belgian, English, Scottish, Dutch, Swiss, German, Irish \\
	East Asian        & 2512  & Korean, Chinese, Japanese                                                 \\ \hline
	Middle Eastern    & 645   & Iranian, Jewish, Lebanese, Turkish                                        \\
	South Asian       & 621   & Bangladeshian, Indian, Pakistani                                          \\
	Southeast Asian   & 457   & Indonesian, Malaysian, Filipino, Thai, Vietnamese                         \\
	Eastern European  & 381   & Eastern European, Russian                                                 \\
	African           & 352   & Moroccan, East African, North African, South African, West African        \\
	Northern European & 250   & Scandinavian                                                              \\
	\hline\hline
    \end{tabular}

\label{tab:cuisine_stat}
\end{table} 

\begin{table}\centering 

\begin{tabular}{r | c | c | c}

                    & Epicurious vs. Allrecipes & Epicurious vs. Menupan & Allrecipes vs. Menupan \\ \hline \hline
North American      & 0.93                      & N/A                    & N/A\\
East Asian          & 0.94                      & 0.79                   & 0.82 \\
Western European    & 0.92                      & 0.88                   & 0.89 \\
Southern European   & 0.93                      & 0.83                   & 0.83 \\
Latin American      & 0.94                      & 0.69                   & 0.74 \\ \hline \hline
African             & 0.89                      & N/A                    & N/A\\
Eastern European    & 0.93                      & N/A                    & N/A\\
Middle Eastern      & 0.87                      & N/A                    & N/A \\
Northern European   & 0.77                      & N/A                    & N/A\\
South Asian         & 0.97                      & N/A                    & N/A\\
Southeast Asian     & 0.92                      & N/A                    & N/A\\

\end{tabular}

\caption[Coherence of cuisines]{The correlation of ingredient usage between
different datasets.  We see that the different datasets broadly agree on what
constitutes a cuisine, at least at a gross level.\label{tab:corr_data}}

\end{table} 

\subsubsection{Size of recipes}\label{sec:size_recipes} 

\begin{table}\centering 
\begin{tabular}{l | l}
\hline
North American & 7.96\\
Western European & 8.03\\
Southern European & 8.86\\
Latin American & 9.38\\
East Asian & 8.96\\
\hline \hline
Northern European & 6.82 \\
Middle Eastern & 8.39\\
Eastern European  & 8.39\\
South Asian & 10.29\\
African & 10.45\\
Southeast Asian & 11.32\\
\hline
\end{tabular}

\caption[Each cuisine's average number of ingredients per recipe]{Average
number of ingredients per recipe for each cuisine.
\label{tab:average_recipe_size}}

\end{table} 

\begin{figure}[b!]\centering 

    \includegraphics[width=0.9\textwidth]{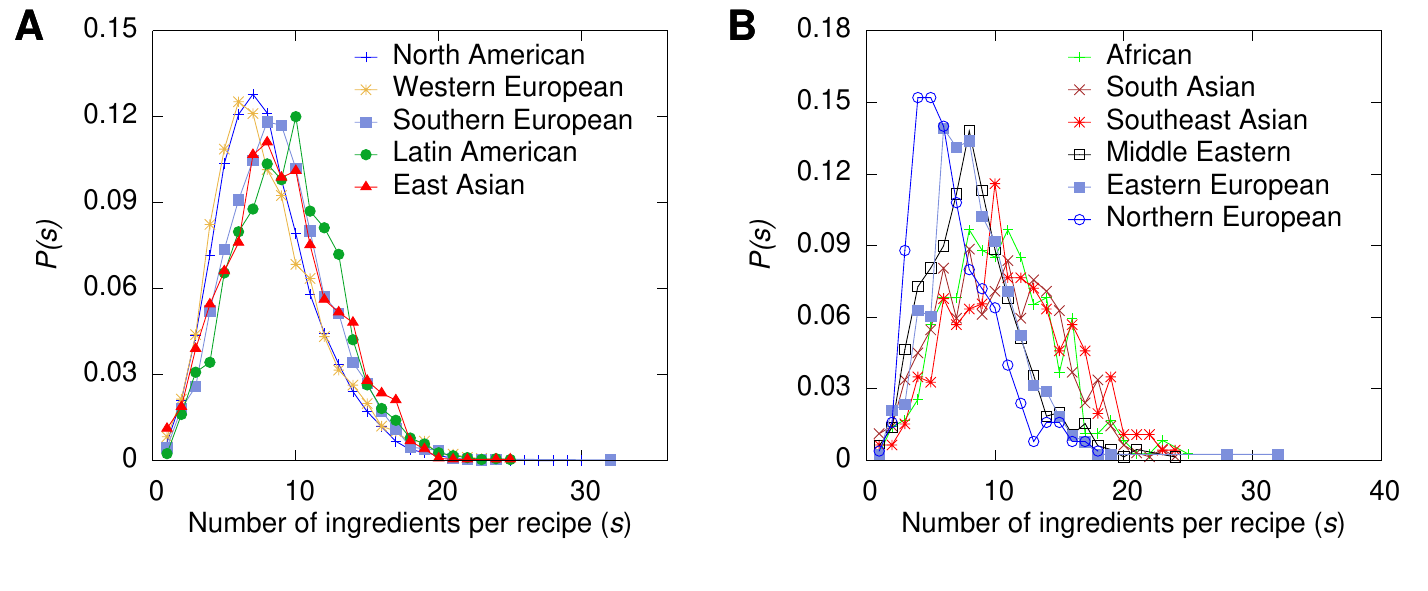}

    \caption[Number of ingredients per recipe]{Number of ingredients per
    recipe. North American and Western European cuisine shows similar
    distribution while other cuisines have slightly more ingredients per
    recipe. \label{fig:recipe_size_dist}}

\end{figure} 

We reports the size of the recipes for each cuisine in
Table~\ref{tab:average_recipe_size}. Overall, the mean number of ingredients
per recipe is smaller than that reported in Kinouchi et
al.~\cite{Skinouchi_culinaryevol_2008}. We believe that it is mainly due to the
different types of data sources. There are various types of recipes: from quick
meals to ones used in sophisticated dishes of expensive restaurants; likewise,
there are also various cookbooks. The number of ingredients may vary a lot
between recipe datasets. If a book focuses on sophisticated, high-level dishes
then it will contain richer set of ingredients per recipe; if a book focuses on
simple home cooking recipes, then the book will contain fewer ingredients per
recipe.  We believe that the online databases are close to the latter; simpler
recipes are likely to dominate the database because anyone can upload their own
recipes. By contrast, we expect that the cookbooks, especially the canonical
ones, contain more sophisticated and polished recipes, which thus are more
likely to contain more ingredients. 

Also, the pattern reported in Kinouchi et
al.~\cite{Skinouchi_culinaryevol_2008} is reversed in our dataset: Western
European cuisine has 8.03 ingredients per recipe while Latin American cuisine
has 9.38 ingredients per recipe.  Therefore, we believe that there is no clear
tendency of the number of ingredients per recipe between Western European and
Latin American cuisine. 

Yet, there seems to be an interesting trend in our dataset that hotter
countries use more ingredients per recipe, probably due to the use of more
herbs and spices~\cite{Sbilling1998,Ssherman2001} or due to more diverse
ecosystems. (6.82 in Northern European vs. 11.31 in Southeast Asian).
Figure~\ref{fig:recipe_size_dist} shows the distribution of recipe size in all
cuisines. 


\subsubsection{Frequency of recipes}\label{sec:freq_recipes} 

\begin{figure}\centering 

    \includegraphics[width=0.5\textwidth]{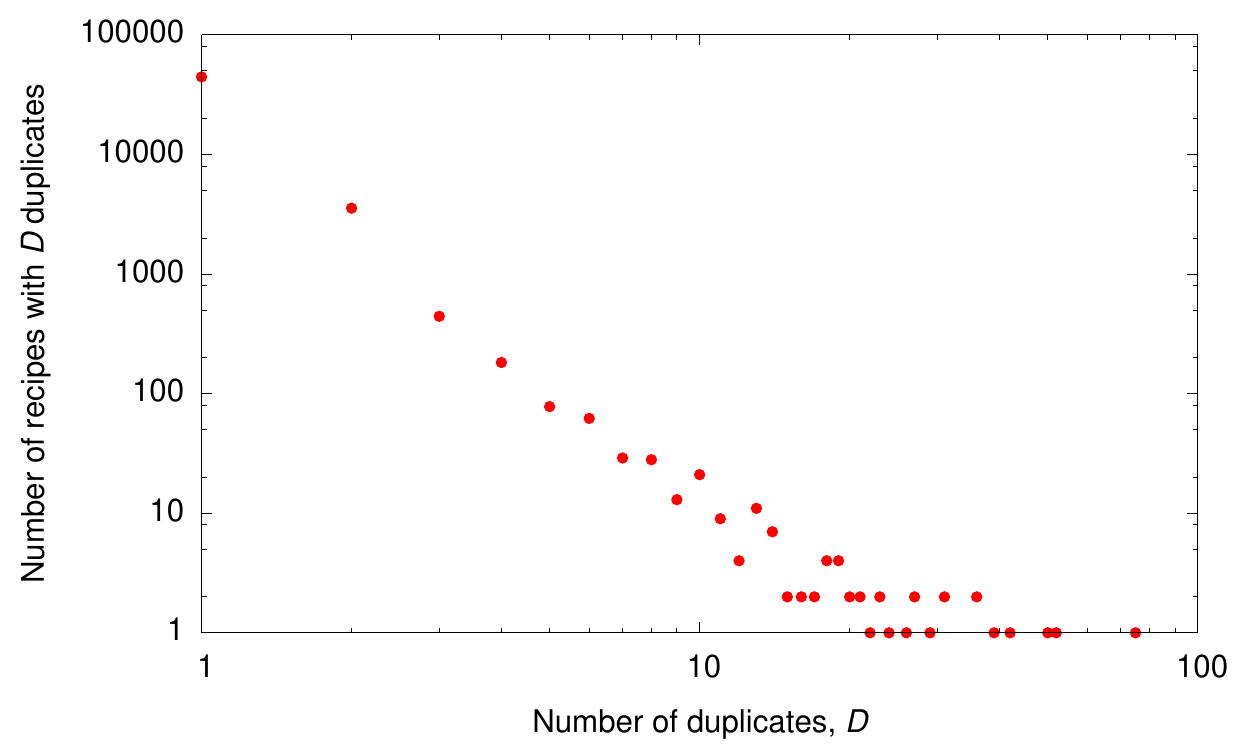}

    \caption[The distribution of duplicated recipes]{If a recipe is very
    popular, the recipe databases will have a tendency to list more variations
    of the recipe. This plot shows that there are many duplicated recipes that
    share the same set of ingredients. The number of duplicates exhibits a
    heavy-tailed distribution. \label{fig:duplicates}}

\end{figure} 

In contrast to previous work~\cite{Skinouchi_culinaryevol_2008} that used
published cookbooks, we use online databases. Although recipes online are
probably less \emph{canonical} than established cookbooks, online databases
allow us to study much larger dataset more easily. Another important benefit
of using online databses is that there is no \emph{real-estate issue} in
contrast to physical cookbooks that should carefully choose what to include.
Adding a slight variation of a recipe costs virtually nothing to the websites
and even enhances the quality of the database. Therefore, one can expect that
online databases capture the frequency of recipes more accurately than
cookbooks.

Certain recipes (e.g. signature recipes of a cuisine) are much more important
than others; They are cooked much more frequently than others.
Figure~\ref{fig:duplicates} shows that there are many duplicated recipes
(possessing identical sets of ingredients), indicating that popularity is
naturally encoded in these datasets. 



\subsection{Number of shared compounds} 

\begin{figure}[b]\centering 

    \fbox{\includegraphics[width=0.4\columnwidth, trim=10 470 0 0,
    clip=true]{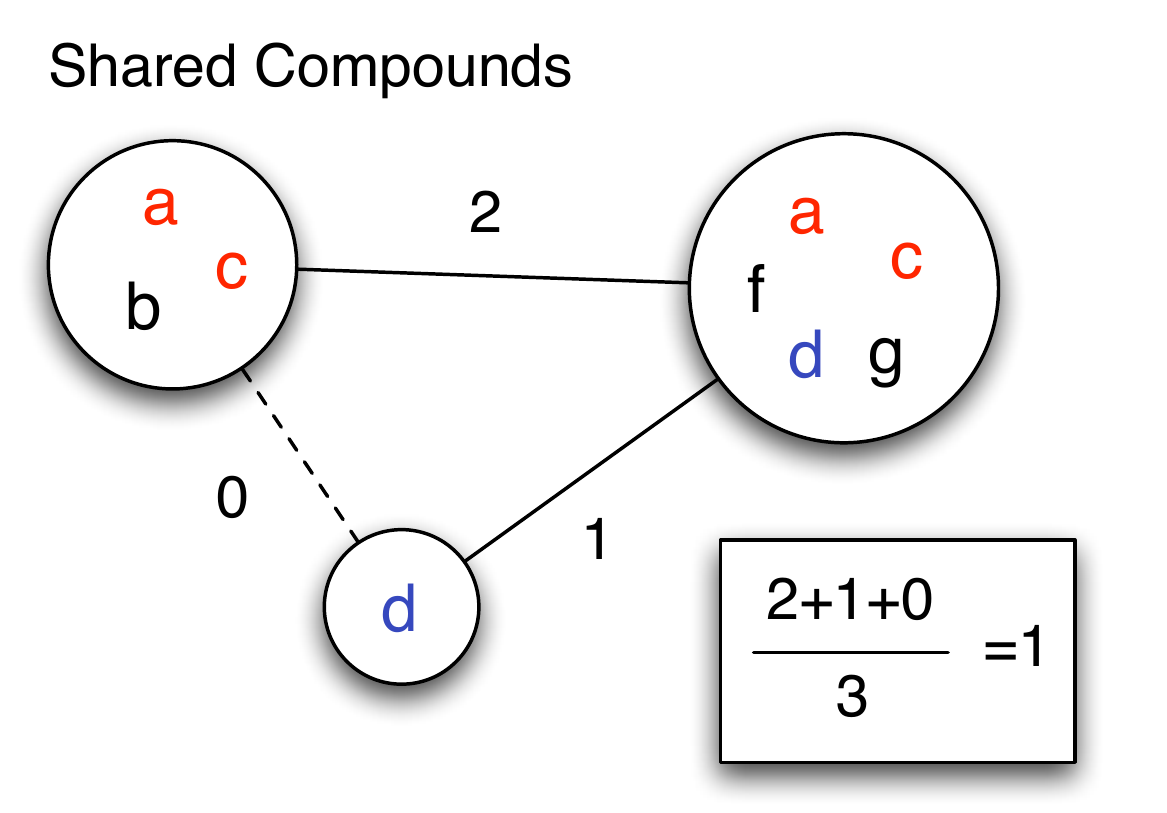}}

    \caption[Measures]{For a recipe with three ingredients, we count the number
    of shared compounds in every possible pair of ingredients, and divide it by
    the number of possible pair of ingredients. } \label{fig:schematic_cartoon}

\end{figure} 

Figure~\ref{fig:schematic_cartoon} explains how to measure the number of shared
compounds in a hypothetical recipe with three ingredients. 


\subsection{Shared compounds hypothesis} 

\subsubsection{Null models} 

\begin{figure}\centering 

    \includegraphics[width=0.7\textwidth]{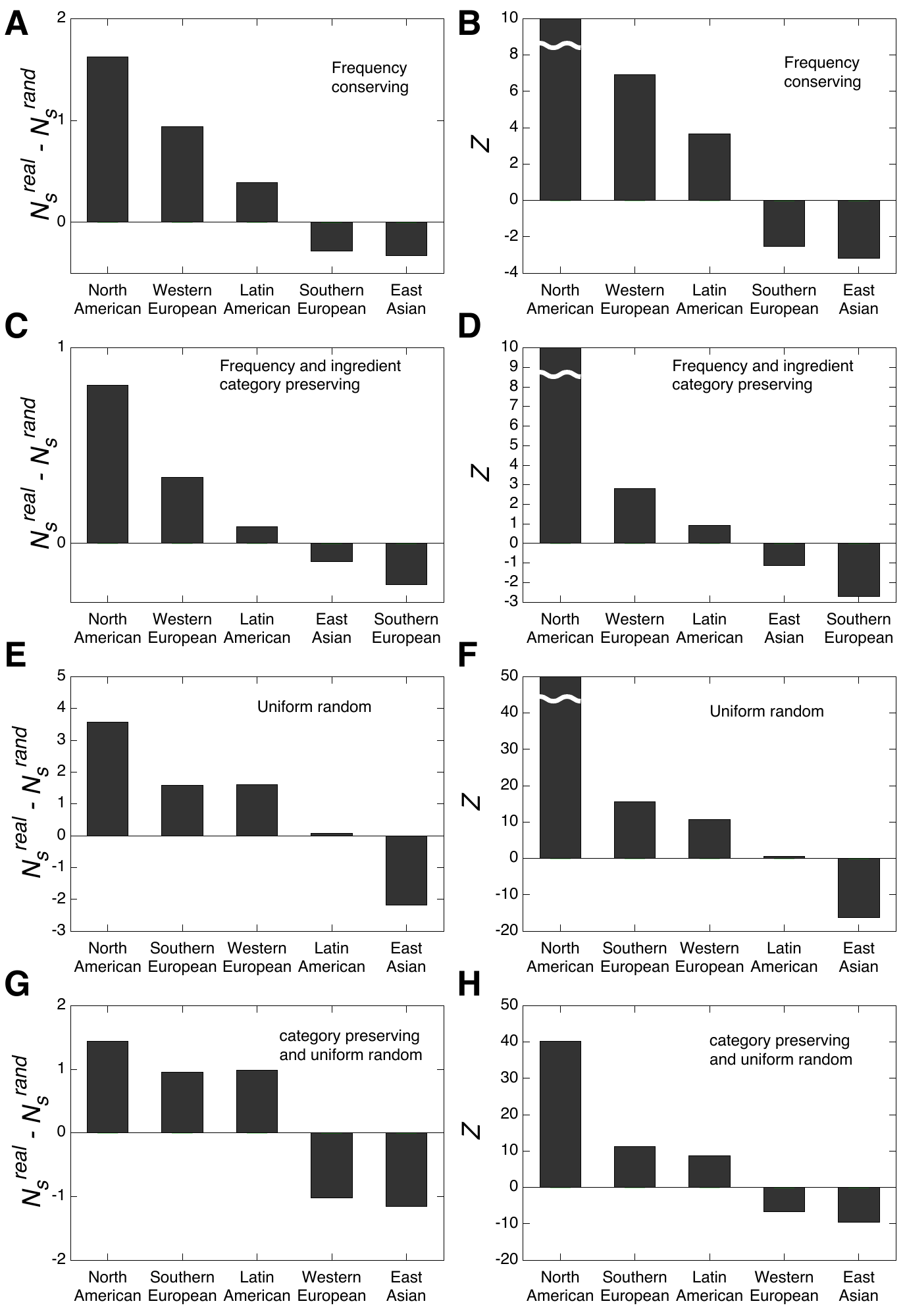} 

    \caption[Null models]{Four different null models. Although the size of the
    discrepancy between cuisines varies greatly, the overall trend is stable.
    \label{fig:null_models}}

\end{figure} 

In order to test the robustness of our findings, we constructed several random
recipe datasets using a series of appropriate null models and compare the mean
number of shared compounds $N_s$ between the real and the randomized recipe
sets. The results of these null models are summarized in
Fig.~\ref{fig:null_models}, each confirming the trends discussed in the paper.
The null models we used are:

\begin{description} 

    \item[(A, B) Frequency-conserving.] Cuisine $c$ uses a set of $n_c$
    ingredients, each with frequency $f_i$. For a given recipe with $N_i$
    ingredients in this cuisine, we pick $N_i$ ingredients randomly from the
    set of all $n_c$ ingredients, according to $f_i$. That is the more
    frequently an ingredient is used, the more likely the ingredient is to be
    picked. It preserves the prevalence of each ingredient. This is the null
    model presented in the main text. 
    
    \item[(C, D) Frequency and ingredient category preserving.]	With this null
    model, we conserve the category (meats, fruits, etc) of each ingredient in
    the recipe, and when sample random ingredients proportional to the
    prevalence. For instance, a random realization of a recipe with beef and
    onion will contain a meat and a vegetable. The probability to pick an
    ingredient is proportional to the prevalence of the ingredient in the
    cuisine. 
    
    \item[(E, F) Uniform random.] We build a random recipe by randomly choosing
    an ingredient that is used at least once in the particular cuisine. Even
    very rare ingredients will frequently appear in random recipes. 

    \item[(G, H) Uniform random, ingredient category preserving.] For each
    recipe, we preserve the category of each ingredient, but not
    considering frequency of ingredients.  

\end{description} 

Although these null models greatly change the frequency and type of ingredients
in the random recipes, North American and East Asian recipes show a robust
pattern: North American recipes always share more flavor compounds than
expected and East Asian recipes always share less flavor compounds than
expected. This, together with the existence of both positive and negative
$N_s^{real} - N_s^{rand}$ in every null model, indicates that the patterns we
find are not due to a poorly selected null models. 

\begin{figure}\centering 

    \includegraphics[width=0.7\textwidth]{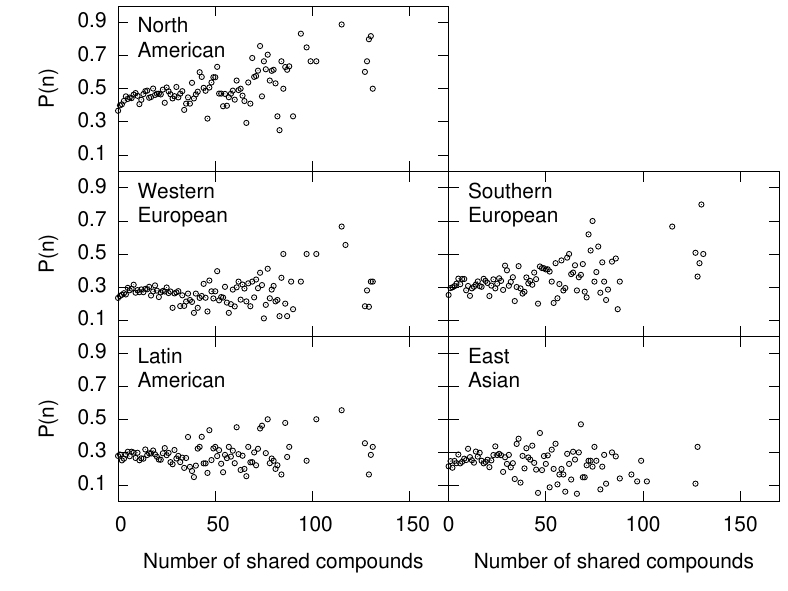}	

    \caption[Shared compounds and usage] {The probability that ingredient pairs
    that share a certain number of compounds also appear in the recipes. We
    enumerate every possible ingredient pair in each cuisine and show the
    fraction of pairs in recipes as a function of the number of shared
    compounds. To reduce noise, we only used data points calculated from more
    than 5 pairs.
    \label{fig:shared_vs_present}}

\end{figure} 

Finally, Fig.~\ref{fig:shared_vs_present} shows the probability that a given
pair with certain number of shared compounds will appear in the recipes,
representing the raw data behind the generalized food-pairing hypothesis
discussed in the text. To reduce noise, we only consider $N_s$ where there are
more than five ingredient pairs.  


\subsubsection{Ingredient contributions} 

To further investigate the contrasting results on the shared compound
hypothesis for different cuisines, we calculate the contribution of each
ingredient and ingredient pair to $\Delta N_s$. Since $N_s(R)$ for a recipe $R$
is defined as
\begin{equation} 
N_s(R) = \frac{2}{n_R(n_R-1)} \sum_{i,j \in R,  i \neq j} \left | C_{i} \cap
C_{j} \right | 
\end{equation}  
(where $n_R$ is the number of ingredients in the recipe $R$), the contribution
from an ingredient pair $(i,j)$ can be calculated as following:
\begin{equation} 
\chi_{ij}^c = \left ( \frac{1}{N_c} \sum_{R \ni i,j} \frac{2}{n_R(n_R-1)} \left
| C_{i} \cap C_{j} \right | \right) -  \left ( \frac{f_{i}f_{j}}{N_c^2}
\frac{2}{\left < n_R \right > (\left <n_R \right >-1)} \left | C_{i} \cap C_{j}
\right | \right ),
\end{equation} 
where $f_i$ indicates the ingredient $i$'s number of occurrences.  Similarly,
the individual contribution can be calculated: 
\begin{equation} 
\chi_{i}^c = \left ( \frac{1}{N_c} \sum_{R \ni i} \frac{2}{n_R (n_R - 1)}
\sum_{j \neq i (j,i \in R)} \left | C_{i} \cap C_{j} \right | \right) - \left (
\frac{2 f_i}{N_c \langle n_R \rangle} \frac{\sum_{j \in c} f_j \left | C_{i}
\cap C_{j} \right |}{\sum_{j \in c} f_j} \right ).
\end{equation} 

We list in Table.~\ref{tab:topcontributors} the top contributors in North
American and East Asian cuisines.

\begin{table}\centering 

\begin{tabular}{| l |c | c ||c|c|}
\hline
& \multicolumn{2}{|c||}{North American} & \multicolumn{2}{|c|}{East Asian} \\
\hline
 & Ingredient $i$ & $\chi_{i}$ & Ingredient $i$ & $\chi_{i}$ \\ 
\hline
\multirow{15}{*}{Positive} & milk & 0.529 & rice       &  0.294 \\ 
& butter         & 0.511                  & red bean   &  0.152 \\ 
& cocoa          & 0.377                  & milk       &  0.055 \\ 
& vanilla        & 0.239                  & green tea  &  0.041 \\ 
& cream          & 0.154                  & butter     &  0.041 \\ 
& cream cheese   & 0.154                  & peanut     &  0.038 \\ 
& egg            & 0.151                  & mung bean  &  0.036 \\ 
& peanut butter  & 0.136                  & egg        &  0.033 \\ 
& strawberry     & 0.106                  & brown rice &  0.031 \\ 
& cheddar cheese & 0.098                  & nut        &  0.024 \\ 
& orange         & 0.095                  & mushroom   &  0.022 \\ 
& lemon          & 0.095                  & orange     &  0.016 \\ 
& coffee         & 0.085                  & soybean    &  0.015 \\ 
& cranberry      & 0.070                  & cinnamon   &  0.014 \\ 
& lime           & 0.065                  & enokidake  &  0.013 \\
\hline
\multirow{15}{*}{Negative} & tomato & -0.168  & beef & -0.2498 \\
& white wine      & -0.0556                   & ginger &  -0.1032 \\
& beef            & -0.0544                   & pork    &  -0.0987 \\
& onion           & -0.0524                   & cayenne &  -0.0686 \\
& chicken         & -0.0498                   & chicken &  -0.0662 \\
& tamarind        & -0.0427                   & onion &  -0.0541  \\
& vinegar         & -0.0396                   & fish &  -0.0458 \\
& pepper          & -0.0356                   & bell pepper &  -0.0414 \\
& pork            & -0.0332                   & roasted sesame seed &-0.0410   \\
& celery          & -0.0329                   & black pepper & -0.0409 \\
& bell pepper     & -0.0306                   & shrimp & -0.0408 \\
& red wine        & -0.0271                   & shiitake & -0.0329 \\
& black pepper    & -0.0248                   & garlic & -0.0302 \\
& parsley         & -0.0217                   & carrot & -0.0261 \\
& parmesan cheese & -0.0197                   & tomato & -0.0246  \\
\hline

\end{tabular}

\caption[Top contributors in North American and East Asian cuisines]{Top 15
(both positive and negative) contributing ingredients to each cuisine.
\label{tab:topcontributors}}

\end{table} 




\clearpage

%
\bibliographystyle{naturemag}

\end{document}